\documentstyle[prl,aps]{revtex}
\def\be{\begin{equation}}
\def\ee{\end{equation}}
\def\bea{\begin{eqnarray}}
\def\eea{\end{eqnarray}}
\def\bar{\begin{array}}
\def\ba{\begin{array}}

\begin{document}

\vspace*{3cm}

\large

{\Large \bf Oscillations of spherical and cylindrical shells}

\hspace{10pt} Serkan Erdin,$^{a)}$ and   Valery L. Pokrovsky $^{a,b)}$

\hspace{10pt} $^{a)}$ Department of Physics, Texas A\&M University,   
College Station  

\hspace{10pt} TX 77843-4242  

\hspace{10pt} $^{b)}$ Landau Institute for Theoretical Physics, Moscow
117940, Russia

\vspace{10pt}

\begin{abstract}
\large
\noindent  We have found  the complete spectrum and eigenstates for
harmonic
oscillations of
ideal spherical and cylindrical shells, both being infinetely thin. The
spectrum of the cylindrical shell has an infinite number of Goldstone
modes corresponding to
folding deformations. This infrared catastrophe is overcome by accounting
for curvature-dependent part of energy. 
\end{abstract}

\pagebreak 


\noindent {\bf INTRODUCTION}

 \ \  \ \ \  The oscillating surfaces have numerous applications in many
areas of physics, spanning from biophysics and chemical physics to high
energy physics. $^{1,2,3}$In this article, we study  the oscillation
spectra of spherical and cylindrical
surfaces, considering them as infinitely thin elastic membranes. We also
calculate their heat capacities. Corresponding calculated values of heat
capacity are valid at sufficiently low temperature $T << \Theta_D$, where
$\Theta_D$ is Debye temperature, otherwise the effects of discreteness
become essential. To our knowledge and big surprise, this problem has not
yet been solved, despite of a vast literature on the subject. $^{4}$ 


\noindent {\bf THE SPECTRUM AND EIGENSTATES OF SPHERICAL AND
CYLINDRICAL SHELLS}

\noindent {\bf  A. Spherical Membrane}

\noindent The Lagrangian of a  spherical membrane reads :
\bea
{\cal L} &=& \frac{\rho}{2} \int R^2 d\Omega \Bigl [ \bigl (
\frac{\partial u_r}{\partial t} \bigr )^2 +  \bigl ( \frac{\partial
u_{\theta}}{\partial t} \bigr)^2 + \bigl ( \frac{\partial 
u_{\phi}}{\partial t} \bigr )^2 \Bigr ] \nonumber \\ 
&-& \frac{\lambda}{2} \int R^2 d \Omega ( U_{\theta\theta} +
U_{\phi \phi} )^2 - \mu \int R^2 d \Omega ( U_{\theta \theta}^2 + 2
U_{\theta \phi}^2 + U_{\phi \phi}^2 )  
\eea
\noindent where $R$ is radius of spherical membrane, $d\Omega =
\sin{\theta}
d\theta d\phi$ and, $\mu$ and $\lambda$ are the Lame coefficients. $u_r$,
$u_\theta$, $u_\phi$ are components of the displacement vector and
$U_{\theta \theta}$, $U_{\phi \phi}$, $U_{\theta \phi}$ are components of
the strain tensor. The latter are determined by relative change of
distance
between two points at a deformation : $d s^{' 2} = d s^2 ( 1 + U_{\alpha
\beta} d x_\alpha d x_\beta )$ and can be expressed in terms of the
displacement vector and its derivatives as follows:  $^{5}$ 
\begin{equation}
U_{\theta \theta} = \frac{1}{R} \frac{\partial u_\theta}{\partial \theta}
+ \frac{u_r}{R} 
\end{equation}
\begin{equation}
U_{\phi \phi} = \frac{1}{R \sin{\theta}} \frac{\partial u_{\phi}}{\partial
\phi} + \frac{u_\theta}{R} \cot{\theta} + \frac{u_r}{R}
\end{equation}
\begin{equation}
2 U_{\theta \phi} = \frac{1}{R} \bigl ( \frac{\partial u_\phi}{\partial
\theta} - u_{\phi} \cot{\theta} \bigr ) + \frac{1}{R \sin{\theta}}
\frac{\partial u_\theta}{\partial \phi}
\end{equation}
\noindent Equations of motions for oscillations with the frequency
$\omega$ read :
\begin{equation}
u_r = {\cal K} \bigl
(\frac{\partial u_\theta}{\partial \theta} + u_{\theta} \cot{\theta} +
\frac{1}{R \sin{\theta}} \frac{\partial u_\phi}{\partial \phi} \bigr)
\end{equation}
\bea
&\Biggl ( \lambda + 2 \mu + 2 ( \lambda + \mu ) {\cal K} \Biggr ) \Bigl (
\frac{\partial^2 u_\theta}{\partial
\theta^2} + \frac{\partial ( u_\theta \cot{\theta} )}{\partial \theta}
\Bigr ) +  \Biggl ( \lambda + \mu +  2 (\lambda + \mu) {\cal K} \Biggr )
\frac{1}{\sin \theta} \frac{\partial^2
u_\phi}{\partial \theta \partial \phi} \nonumber \\
&- \Biggl (\lambda + 3 \mu +  2 (\lambda + \mu) {\cal K} \Biggr )
\frac{\cot
\theta}{\sin \theta}
\frac{\partial u_\phi}{\partial \theta} + \frac{\mu}{\sin^2 \theta}
\frac{\partial^2 u_\theta}{\partial \phi^2} 
+  (\rho \omega^2 R^2 + 2\mu) u_\theta = 0
\eea 
\bea
&\Biggl ( \lambda +  \mu +  2 (\lambda + \mu) {\cal K} \Biggr ) \Bigl (
\frac{1}{\sin
\theta} \frac{\partial^2
u_\theta}{\partial \theta \partial \phi} \Bigr ) 
+ \Biggl ( \lambda + 3 \mu +  2 (\lambda + \mu) {\cal K} \Biggr ) \Bigl (
\frac{\cot \theta}{\sin \theta}
\frac{\partial u_\theta}{\partial \phi} \Bigr )  \nonumber \\
&+ \Biggl ( \lambda + 2 \mu +  2 (\lambda + \mu) {\cal K} \Biggr ) \Bigl
(
\frac{1}{\sin^2 \theta}
\frac{\partial^2 u_\theta}{\partial \phi^2} \Bigr ) + \mu \Bigl (
\frac{\partial^2 u_\phi}{\partial \theta^2} + \frac{\partial ( u_\phi
\cot{\theta} )}{\partial \theta} \Bigr ) \nonumber  \\ 
&+ ( \rho \omega^2 R^2 + 2\mu ) u_\phi = 0  
\eea
\noindent where ${\cal K} =  \frac{2 (\lambda + \mu)}{\rho \omega^2 R^2 -
4(\lambda + \mu)}$.
\noindent There exists a symmetric solution (the "breathing" mode),
$u_\theta =
u_\phi = 0$ , $u_r = const \not = 0$. The frequency of this solution is
$\omega_s = \sqrt{\frac{ 4(\lambda + \mu)}{\rho R^2}}$. From other
solutions, we first consider those independent on $\phi$. There exist
three branches of
such oscillations. The first two ("longitudinal" modes) correspond to
$u_\phi =
0$, $u_\theta \not =
0$, $u_r \not = 0$. The frequency of these oscillations are labeled by an
integer $l$ :
\bea
{\omega}^2_{\pm, l} &=& \frac{[ 4\lambda + 2\mu + (\lambda + 2\mu) l(l +
1)]}{2 \rho R^2} \nonumber \\
&\pm& \frac{\sqrt{ ( 4\lambda + 2\mu + (\lambda + 2\mu) l(l +   
1))^2 - 16 \mu (\lambda + \mu)(l(l +1) - 2)}}{2 \rho R^2} 
\eea
\noindent In the third branch (a "shear" mode) $u_\theta = 0$ , $u_r = 0$
and $u_\phi \not
= 0$. Its
frequencies are 
\begin{equation}
{\omega}^2_{t, l} = \frac{\mu}{\rho R^2} (l(l + 1) - 2)
\end{equation}
\noindent where $l = 2,3, ...$ Explicit expressions for the displacements  for the
longitudinal  branches are: $^{6}$ 
\be
u_\phi = 0  \hspace{1cm}   u_\theta = C \frac{d P_l (\cos \theta )}{d
\theta}
\ee
\be 
u_r = - \frac{C (\lambda + \mu)}{\rho \omega^2 R^2 - 4(\lambda + \mu)}
l(l + 1) P_l (\cos \theta) 
\ee
\noindent where $C$ is constant amplitudes and $P_l (\cos
\theta)$ are
the
Legendre
polynomials.
\noindent The displacements in the shear branch solution read :   
\be
u_r = u_\theta = 0 \hspace{1cm}    u_\phi = C \frac{d P_l (\cos
\theta)}{ d \theta} 
\ee

In order to get eigenstates for $m \not = 0$, we  employ
an infinitesimal rotation around $x$
axis which turns an arbitrary vector $\vec r$ into :
\be
\vec r^{'} = O( \omega ) \vec{r} = 1 + \omega g_x \vec{r}
\ee
 \noindent Here $-i g_x$ is the generator of the rotation around $x$ axis.
$^{7}$ It generates a transformation of the displacement field $\vec
u(\vec r)$.
\be
\vec u^{'}(\vec r) = O(\omega) \vec u \bigl (O^{-1} (\omega) \vec r \bigr
) = \vec u (\vec r) - \omega g_x \vec u (\vec r ) + \omega (g_x \vec
r).\nabla u (\vec r) ...
\ee
 \noindent If $u_0 (\vec r)$ was a solution of the system (5-7), then
$\vec u^{'}
(\vec r)$ is also a solution. The operation we performed to find new
solutions is readily recognizable as the Lee translation introduced in
theory
of fiber bundles. $^{8}$  

After
a simple transformation, we find an operator ${\cal G}$, which, acting on
a solution fixed by two components
$\bordermatrix{&  \cr
               & u_\theta \cr
               & u_\phi   \cr}$ ,  generates another solution with the same
frequency.  
\bea
{\cal G} = \bordermatrix{&                          \cr
& \sin \phi \frac{\partial}{\partial \theta} + \cot \theta \sin \phi
\frac{\partial}{\partial \phi}    & - \frac{\cos \phi}{\sin \theta} \cr
& \frac{\cos \phi}{\sin \theta}    & \sin \phi \frac{\partial}{\partial
\theta} + \cot \theta \sin \phi
\frac{\partial}{\partial \phi}  \cr}
\eea
\noindent The operator ${\cal G}$ is the analog of the operator of total
angular
momentum ${\cal J}$ in quantum mechanics, but it is reduced to a subspace
of tangential components of the displacement vector $\vec u$.
The Lagrangian (1) and equations of motion (6-7) are invariant with
respect
to the transformation (15). Acting by the operator ${\cal G}$ onto
solutions
with $m = 0$ (10-12), we generate solutions with all possible values of
$m$ .
They belong to the same eigen-frequencies as the  initial
solutions
with $m = 0$. The displacements  for solutions belonging to the
longitudinal branches with $m \not = 0$ read : 
\be
u_{\theta}^{(m)} (x) = e^{i m \phi} i^{m} \Biggl [ m x (1 - x^2 )^{\frac{m
-
1}{2}} P_{l}^{(m)} (x) - (1 - x^2)^{\frac{m + 1}{2}} P_{l}^{(m + 1)} (x)
\Biggr ] 
\ee
\be 
u_{\phi}^{(m)} (x) = - e^{ i m \phi} i^{m + 1} m (1 - x^2)^{\frac{m -
1}{2}}
P_{l}^{(m)} (x)
\ee
\noindent The displacements for oscillations of the shear branch with $m
\not = 0$ are : 
\be
u_{\theta}^{(m)} (x) = - e^{ i m \phi} i^{m + 1} m (1 - x^2)^{\frac{m -
1}{2}} 
P_{l}^{(m)} (x)   
\ee
\be
u_{\phi}^{(m)} (x) = e^{i m \phi} i^{m} \Biggl [ m x (1 - x^2 )^{\frac{m -
1}{2}} P_{l}^{(m)} (x) - (1 - x^2)^{\frac{m + 1}{2}} P_{l}^{(m + 1)} (x)
\Biggr ] 
\ee

Given the known frequency spectrum $\omega_j$, the thermal energy can be
calculated as $U = \sum_{j} \frac{\hbar \omega_j}{exp \frac{\hbar
\omega_j}{T} -1}$ and the heat capacity can be calculated as $C = \bigl (
\frac{\partial U}{\partial T} \bigr )_V$.

As always, low-lying rotational frequencies $\omega_{k}^{(rot)} =
\frac{\hbar k ( k + 1 )}{2 M R^2}$ are smaller than low-lying oscillatory
frequencies by a factor $\sim \sqrt{\frac{m}{M}} \frac{a}{R}$, where $a$
is characteristic distances between atoms ( discreetness effect ) and $m$
is the mass of electron. That means that at temperature $T >> T_R =
\frac{\hbar^2}{M R^2}$, the rotational degrees of freedom become
substantially classical until $k \leq \sqrt{\frac{T}{T_R}}$.Their
contribution to the heat capacity is $C_R = \frac{T}{T_R} \zeta(3)$.
Elastic oscillations remain frozen till a larger temperature $T_0
\sim \frac{\hbar}{R} \sqrt{\frac{\mu}{\rho}} >> T_R$.

\noindent {\bf B. Cylindrical shells}

In order to find the spectrum and corresponding eigenstates of
a cylindrical
membrane, we follow the same steps as we did for spherical shell.
For cylindrical case, the components of strain tensor in terms of
cylindrical coordinates $r$ , $\phi$, $z$ read: $^{5}$ 
\be
U_{\phi \phi} = \frac{1}{R} \frac{\partial u_\phi}{\partial \phi} +
\frac{u_r}{R}, \hspace {30pt} U_{zz} = \frac{\partial u_z}{\partial z},
\hspace{30pt} 2
U_{\phi z } =
\frac{1}{R} \frac{\partial u_z}{\partial \phi} +
\frac{\partial u_\phi}{\partial z}
\ee
\noindent The Lagrangian reads :
\bea
{\cal L} &=& \frac{\rho}{2} \int R d \phi d z \Biggl [ \Bigl (
\frac{\partial u_r}{\partial t} \Bigr )^2 + \Bigl (
\frac{\partial u_\phi}{\partial t} \Bigr )^2 +\Bigl (
\frac{\partial u_z}{\partial t} \Bigr )^2 \Biggr ] \nonumber \\ 
&-& \frac{\lambda}{2} \int R d \phi d z ( U_{zz} + U_{\phi \phi})^2 - \mu
\int R d \phi d z ( U_{zz}^{2} + 2 U_{z \phi}^{2} + U_{\phi \phi}^{2} )
\eea
\noindent where $R$ is the radius of the cylinder. Equations of motions
are :
\be 
u_r = (\lambda + 2 \mu) \tilde {\cal K}
\frac{\partial u_\phi}{\partial \phi} + \tilde {\cal K} R \lambda
\frac{\partial u_z}{\partial z}
\ee
\be
 u_\phi = -  ( \lambda + 2
\mu) \tilde {\cal K} \frac{\partial^2
u_\phi}{\partial \phi^2} 
-  ( \lambda + \mu) \tilde {\cal K}  \frac{\partial^2 u_z}{\partial z
\partial \phi}  +
\frac{\mu}{\rho \omega^2 R^2} \biggl ( ( \lambda + 2
\mu) \tilde {\cal K} R \frac{\partial^2 u_z}{\partial z
\partial \phi} - R^2 \frac{\partial^2 u_\phi}{\partial z^2} \biggr )
\ee
\bea
 u_z &=& -  (\lambda + \mu) \tilde {\cal K} R
\frac{\partial^2 u_\phi}{\partial z \partial \phi} - (\lambda + 2\mu)
\tilde {\cal K}  R^2
\frac{\partial^2 u_z}{\partial z^2}  \nonumber  \\   
&+& \frac{\mu}{\rho \omega^2 R^2} \biggl ( (\lambda + 2 \mu) \tilde {\cal
K}  R \frac{\partial^2 u_\phi}{\partial z \partial \phi} +  4  (\lambda +
\mu )
\tilde {\cal K} R^2 \frac{\partial^2 u_z}{\partial z^2} -
\frac{\partial^2 u_z}{\partial \phi^2} \biggr )
\eea
\noindent where $\tilde {\cal K} = \frac{1}{\rho R^2 \omega^2 - ( \lambda
+ 2
\mu)}$.
\noindent Specifying the $z$ dependence of all displacement components as
a
plane wave $e^{i k z}$, and their angular dependence as $e^{i m \phi}$,
we find a system of linear equations for the amplitudes $u_r$, $u_\theta$
, $u_\phi$. Its secular equation reads :
\bea
( \omega^2 - \omega_{s}^{2}) \Biggl ( \omega^6 - \biggl ( \omega_{t}^{2} (
R^2 k^2 + m^2 ) + \omega_{s}^{2} (R^2 k^2 + m^2 + 1) \biggr ) \omega^4 \nonumber \\
+ \biggl ( \omega_{t}^{2} \omega_{s}^{2} \Bigl ( ( R^2 k^2 + m^2 )^2 + 5
R^2 k^2 + m^2 \Bigr ) - 4 \omega_{t}^{4} R^2 k^2 \biggr ) \omega^2
\nonumber \\
+ 4 \omega_{t}^{4} ( \omega_{t}^{2} - \omega_{s}^{2} ) R^4 k^4 \Biggr ) =
0
\eea
\noindent where $\omega_{s}^{2} = \frac{\lambda + 2\mu}{\rho R^2}$ and
$\omega_{t}^{2} = \frac{\mu}{\rho R^2}$. One of its solutions (symmetric)
is $\omega^2 = \omega_{s}^{2}$. For $k = 0$, the remaining cubic equation
can be solved explicitly for arbitrary $m$. The three solutions are :
\be 
\omega^2 = 0 , \hspace{1cm} \omega^2 = (m^2 + 1) \omega_{s}^2 ,
\hspace{1cm} \omega^2 = m^2
\omega_{t}^{2}
\ee
\noindent The first of them  implies that there exists infinite number of
Goldstone modes
corresponding to the folding deformations. For $m = 0$, the cubic
equation gives the following solutions :
\be
\omega^2 = k^2 R^2 \omega_{t}^{2}
\ee
\be 
\omega^2 = \frac{\omega_{s}^{2}(
R^2 k^2 + 1 )}{2} \pm \frac{\sqrt{(\omega_
{s}^{2} (R^2 k^2 + 1))^2 - 16 \omega_{t}^{2} ( \omega_{s}^{2} -
\omega_{t}^{2} ) R^2 k^2}}{2}
\ee

  The physical reason of the peculiarity in the spectrum of cylindrical
shell is an opportunity to fold a cylinder conserving its straight-lines
and its cross-section contour length. Such a deformation is strainless.
The infinite number of folding modes leads to divergence of the heat
capacity at any temperature. This difficulty can be avoided by
introducing terms with higher derivatives into the Lagrangian.
Geometrically, it means that the Gaussian curvature and the mean curvature
must be included. In more physical or chemical terms it means that not
only the variation of distances between atoms, but also variation of
valence angles must be accounted for. Below we write down explicit
formulae for the Gaussian curvature $K$ and the mean curvature $M$ for cylindrical membranes in linear approximation in displacements.
\be
K = - \frac{1}{R} \frac{\partial^2 u_r}{\partial z^2}
\ee
\be
M =  \frac{1}{R} - \frac{u_r}{R^2} - \frac{1}{R^2} \frac{\partial^2
u_r}{\partial \theta^2} - \frac{\partial^2 u_r}{\partial z^2}
\ee
\noindent The term in Lagrangian depending on curvature reads : 
\be
\Delta L = \int \Biggl [ \frac{\nu}{2} (K - K_0)^2 + \frac{\theta}{2} (M -
M_0)^2 \Biggr ] d^2 S 
\ee
\noindent where $K_0 = 0$ and $M_0 = \frac{1}{R}$ are the values of $K$
and $M$ for undeformed cylinder.
With this term in Lagrangian we find the gaps $\Delta_m$ in the spectrum
corresponding to the former folding modes:
\be
\Delta_m = \sqrt{\frac{\theta}{\rho R^4}} (m^2 - 1)
\ee
\noindent Note that the gaps in spectrum due to curvature effect are
propotional to $R^{-2}$ and, therefore, are small in comparison to
gaps due to strain effects ($\propto R^{-1}$). By perturbation method it
is possible to find small $k$ -
dependent corrections to these frequencies :
 $\delta \omega^{2} = \frac{\theta}{\rho R^4} [ 2 k^2 R^2 ( m^2 -1
)] - \frac{(\omega_s^{2} - 2 \omega_{t}^{2}) \omega_{t}^{2} (k
R)^2}{\Delta_{m}^{2} - \omega_{t}^{2} m^2}$  or $\omega_{m}^{2} (k) =
\Delta_{m}^{2}  + V_{m}^{2} k^2$, where $V_{m}^{2} = \frac{\theta}{\rho
R^4} [ 2 R^2 ( m^2 -1)] - \frac{(\omega_s^{2} - 2 \omega_{t}^{2})
\omega_{t}^{2} (R)^2}{\Delta_{m}^{2} - \omega_{t}^{2} m^2}$.
A  clear tendency is that the band becomes more flat with the growth of
$m$.

For cylindrical shell, we can write the thermal energy in the following
form : 
\be
U = \int\limits_{0}^{\infty} \frac{d k}{2 \pi}
\sum_{m=2}^{\infty} \frac{\hbar \omega_m (k) 
}{exp \frac{\hbar \omega_m (k)}{T} - 1}
\ee
 \noindent  In the temperature interval $\hbar \Delta_2 << T << \hbar
\omega_{t}$, we can
neglect all " elastic branches" and substitute the summation over the
"valence" branches by integration. It leads to the energy : 
\be
U = \frac{0.601}{\hbar^2} \sqrt{\frac{1}{\frac{\theta}{\rho R^2}
(\omega_s^{2} - 2
\omega_t^{2} )}} T^3
\ee
\noindent and the heat capacity  :
\be
C_v = \frac{1.803}{\hbar^2} \sqrt{\frac{1}{\frac{\theta}{\rho R^2}
(\omega_s^{2} - 2
\omega_t^{2} )}} T^2
\ee
\noindent For $T << \hbar \Delta_2$, we can neglect the sum over $m$,
since only $\Delta_2$ and $V_2$ contribute to the energy. The
internal energy becomes exponentially small  :
 \be
U = \sqrt{2 \pi  \hbar T} \frac{\Delta_2^{\frac{3}{2}}}{V_2}
e^{-\frac{\hbar \Delta_2}{T}}
\ee
\noindent and the heat capacity :
\be
C_v = \frac{\sqrt{2 \pi \hbar^3}}{T^{\frac{5}{2}}}
\frac{\Delta_2^{\frac{5}{2}}}{V_2} e^{-\frac{\hbar \Delta_2}{T}}
\ee

\noindent {\bf  CONCLUSIONS}
 
We have found complete  spectrum of oscillations for spherical and
cylindrical membranes and corresponding eigenstates.  We have showed that
cylindrical shell
has an infinite number of Goldstone modes due to folding deformations
which do not change the elastic energy of shell. We have also showed that
this catastrophe can be avoided by introducing curvature-dependent terms 
to the Hamiltonian. In addition, heat capacities of spherical and
cylindrical
shells are calculated. They can be measured experimentally as an
additional contribution to
the heat capacity of a solvent in a diluted solution of long molecules.  

\noindent{\bf ACKNOWLEDGMENTS}

This work was inspired by discussion of heat capacity measurements for
buckyballs and nanotubes with Vladimir Kopylov.

\pagebreak

\noindent $^1$ Stanislas Leibler, in  {\it
Statistical
Mechanics of Membranes and Surfaces}, edited by D.R.Nelson, T.Piran,
S.Weinberg, Vol. 5(World Scientific, Singapore, 1989), p.45

\noindent $^2$ J.M. Park and T.C. Lubensky, J.Phys.{\bf 16}, 493 (1996);
E.I. Kats, V.V. Lebedev and A.R. Muratov, JETP Letters {\bf 63}, 216
(1996); J.F. Weather, Nucl. Phys.{\bf 458}, 671 (1996)

\noindent $^3$ {\it Science of Fullerenes and
Carbon
Nanotubes}, edited by M.S. Dresselhaus ( Acedemic, New York, 1996 )

\noindent $^4$ P.M. Naghdi, in  {\it Mechanics of Solids }, edited
by C.
Truesdell, Vol.2 (Springer-Verlag, Berlin, 1972), p.425
 
\noindent $^5$ L.D. Landau and E.M. Lifshitz ,{\it Theory of Elasticity}
(Pergenon, New York, 1970)

\noindent $^6$ W. Soedel , {\it Vibrations of Shells and Plates } (Marcel
Dekker, New York, 1993)

\noindent $^7$ {\it Group Theory in Physics: Introduction}, edited by J.F.
Cornwel (Academic, New York, 1997)

\noindent $^8$ B.F. Schutz, {\it Geometrical Methods of Mathematical
Physics} ( Cambridge University Press, Cambridge, 1980 )



\end{document}